\documentclass[10pt,twocolumn,a4paper]{esaAI}
\usepackage{tabularx}
\title{Shaping Rewards, Shaping Routes: On Multi-Agent Deep Q-Networks for Routing in Satellite Constellation Networks}

\def\authorEmail{manuel.roth@dlr.de}

\author[1]{Manuel M. H. Roth\thanks{Corresponding author. E-Mail: \authorEmail}}
\author[1]{Anupama Hegde}
\author[2]{Thomas Delamotte}
\author[2]{Andreas Knopp}
\affil[1]{Institute of Communications and Navigation, German Aerospace Center (DLR), Oberpfaffenhofen, Germany}
\affil[2]{Chair of Signal Processing, University of the Bundeswehr Munich, Neubiberg, Germany}

\usepackage{url}
\usepackage{calligra}
\usepackage[nolist, nohyperlinks]{acronym}
\begin{acronym}
    \acro{GEO}{Geostationary Orbit}
    \acro{MEO}{Medium Earth Orbit}
    \acro{LEO}{Low Earth Orbit}
    \acro{ISL}[ISL]{Intersatellite Link}
    \acro{ESL}{Earth-Satellite Link}
    \acro{SCN}{Satellite Constellation Network}
    \acro{NTN}{Non-Terrestrial Network}
    \acro{TN}{Terrestrial Network}
    \acro{RTT}{Round-Trip Time}
    \acro{UT}{User Terminal}
    \acro{GW}{Gateway station}
    \acro{USN}{User Satellite Node}
    \acro{UTN}{User Terrestrial Node}
    \acro{ACM}{Adaptive Coding \& Modulation}
    \acro{VCM}{Variable Coding \& Modulation}
    \acro{LCT}{Laser Communication Terminal}
    \acro{RF}{Radio Frequency}
    \acro{KPI}{Key Performance Indicator}
    \acro{QoS}{Quality of Service}
    \acro{QoE}{Quality of Experience}
    \acro{SLA}{Service Level Agreement}
    \acro{SLR}{Service Level Requirement}
    \acro{SPF}{Shortest Path First}
    \acro{GEVR}{God's Eye View Routing}
    \acro{SSPF}{Source-routed Shortest Path First}
    \acro{SR}[SR]{Segment Routing}
    \acro{ELB}{Explicit Load Balancing}
    \acro{DRA}{Datagram Routing Algorithm}
    \acro{TLR}{Traffic-Light-based intelligent Routing strategy}
    \acro{IDLB}{Independent Distributed Load-balanced routing}
    \acro{NHCN}{Next-Hop Cluster Node}
    \acro{SPT}{Shortest Path Tree}
    \acro{BFS}{Breadth-First Search}
    \acro{SDN}[SDN]{Software Defined Networking}
    \acro{CPP}[CPP]{Controller Placement Problem}
    \acro{IP}{Internet Protocol}
    \acro{MPLS}{Multiprotocol Label Switching}
    \acro{GBR}{Guaranteed Bit Rate}
    \acro{MTU}{Maximum Transmission Unit}
    \acro{TCP}{Transmission Control Protocol}
    \acro{VoIP}{Voice over IP}
    \acro{PIM}{Protocol Independent Multicast}
    \acro{PIM-SM}{PIM Sparse Mode}
    \acro{PIM-DM}{PIM Dense Mode}
    \acro{IGMP}{Internet Group Management Protocol}
    \acro{RPF}{Reverse-Path Forwarding}
    \acro{RP}{Rendezvous Point}
    \acro{ML}{Machine Learning}
    \acro{RL}{Reinforcement Learning}
    \acro{MDP}{Markov Decision Process}
    \acroplural{MDP}[MDPs]{Markov Decision Processes}
    \acro{DRL}{Deep Reinforcement Learning}
    \acro{DNN}{Deep Neural Network}
    \acro{DQN}{Deep Q-Network}
    \acro{DDQN}{Double Deep Q-Network}
    \acro{MADRL}{Multi-Agent DRL}
    \acro{CDRL}{Centralized DRL}
    \acro{FD-MADRL}{Fully-Distributed Multi-Agent DRL}
    \acro{GNN}{Graph Neural Network}
    \acro{POMDP}{Partially Observable MDP}
    \acro{Dec-POMDP}{Decentralized Partially Observable MDP}
    \acro{CL-DC}{Centralized Learning - Decentralized Control}
    \acro{MADDPG}{Multi-Agent Deep Deterministic Policy Gradient}
    \acro{DDPG}{Deep Deterministic Policy Gradient}
\end{acronym}

\begin{document}

\makeCustomtitle

\begin{abstract}
Effective routing in satellite mega-constellations has become crucial to facilitate the handling of increasing traffic loads, more complex network architectures, as well as the integration into 6G networks. To enhance adaptability as well as robustness to unpredictable traffic demands, and to solve dynamic routing environments efficiently, machine learning-based solutions are being considered.
For network control problems, such as optimizing packet forwarding decisions according to Quality of Service requirements and maintaining network stability, deep reinforcement learning techniques have demonstrated promising results.
For this reason, we investigate the viability of multi-agent deep Q-networks for routing in satellite constellation networks. We focus specifically on reward shaping and quantifying training convergence for joint optimization of latency and load balancing in static and dynamic scenarios. To address identified drawbacks, we propose a novel hybrid solution based on centralized learning and decentralized control.
\end{abstract}

\section{Introduction}
As larger and more complex \ac{LEO} \acp{SCN} with \acp{ISL} are proposed, effective routing for \acp{NTN} is becoming an increasingly important topic. To support broadband traffic, as well as the integration with terrestrial networks, novel and specifically tailored routing and network management techniques are required.
Considerations for 6G networks include an adaptive, flexible NTN-integration, which highlights the need for new approaches to interact seamlessly with the AI-native network of networks \cite{muscinelli_overview_2022}.
The optimization goal at hand is finding the "best" routing configuration in a highly dynamic, physically large network with non-uniformly distributed traffic and its \ac{QoS} requirements.
In this context, \ac{DRL}-based schemes have been identified to be particularly promising optimization methods for such network control problems \cite{zhang_deep_2019}.
The approaches combine the ability of reinforcement learning to solve control problems effectively with the capacity of \acp{DNN} to approximate complex, high-dimensional functions.
In this investigation, we look into the viability of such \ac{DRL} schemes based on \acp{DQN} for \acp{SCN}.
Various \ac{FD-MADRL} approaches have been proposed for SCNs in the literature \cite{you_toward_2022, soret_q-learning_2023}.
The idea is that each satellite acts as an individual agent and makes decisions based on its local observations.
However, a careful design and reward shaping is required to avoid loops and to find coherent paths.
Learning coordinated behaviors can pose a challenge due to the complex interactions which can arise from the policies of individual agents.
Utilizing \ac{CDRL} techniques instead, enables a larger scope and thus facilitates the establishment of effective end-to-end routes.
But, centralization is difficult due to the physical size of the network.
To improve scalability, hierarchical strategies have been proposed \cite{ali_hierarchical_2020}.
A path is constructed by the assistance of different controllers (or group leaders) at different hierarchical levels.

We base our investigation on these proposed approaches. The main contribution of this work is the analysis of the \ac{FD-MADRL} routing approach in different SCN scenarios, focusing on the joint optimization of latency and load balancing.
Drawing from the results, we propose a hybrid routing solution based on centralized learning and decentralized control.

\subsection{Routing in Satellite Constellation Networks}
\acp{SCN} differ from terrestrial networks in several aspects. Most importantly, there are frequent \ac{ISL} and \ac{ESL} handover events due to the dynamic network topology. Furthermore, traffic requirements are non-uniformly distributed. As the constellation moves relative to Earth, hot spots are geographical rather than topological. Moreover, the on-board processing is typically limited. For this investigation, we assume four ISLs on each satellite as in \cite{roth_distributed_2022}.
To evaluate the performance of a routing scheme, various \ac{QoS} metrics can be considered, e.g. packet dropping rate, latency, throughput, and link utilization \cite{roth_distributed_2022}.

\subsection{Reinforcement Learning Architectures}
\acp{MDP} provide a formal, mathematically idealized framework for modeling reinforcement learning problems \cite{sutton_reinforcement_2018}. In MDPs, an agent interacts at time step $t$ with an environment by choosing an action $a_t$ from a set of possible actions $\calligra{A}$. The interaction results in a new state $s_{t+1}$, and a reward $r_t$ which the agent seeks to maximize.
This results in a trade-off: agents must balance exploiting known rewards through greedy selection versus exploring unknown states and actions with potentially better long-term rewards.
By decomposing the problem into immediate plus discounted future rewards using the discount factor $\gamma$, we can express the optimization by simpler, recursive sub-problems \cite{sutton_reinforcement_2018}. The estimated viability of each action for a given state, the Q-value, can thus be described by:
\begin{equation}
    Q(s_t, a_t) = r_t + \gamma \max_{a_{t+1}} Q(s_{t+1}, a_{t+1})
\end{equation}
Building on this foundation, \acp{DQN} utilize \acp{DNN} to approximate the Q-function in complex routing environments.
During training, agents interact with the environment and store their experiences in replay buffers for increased stability.
While CDRL is based on a singular agent, in FD-MADRL, each satellite is assumed to act as an independent agent with its own model and only partial knowledge of the environment.
While this is well-suited for dynamic environments, in which changes and disruptions have to be handled locally and quickly, it also introduces additional complexity.
The end-to-end coherence and stability becomes questionable, the agents have to learn the behavior of other agents as well.
Complex and unexpected interactions between agents can emerge.

For FD-MADRL, the state space consists of information about the adjacent link loads, the previous hop and the destination.
The action space emcompasses the available ISLs, so the available next hops.

\begin{figure}[t]
    \centering
    \includegraphics[width=.95\columnwidth]{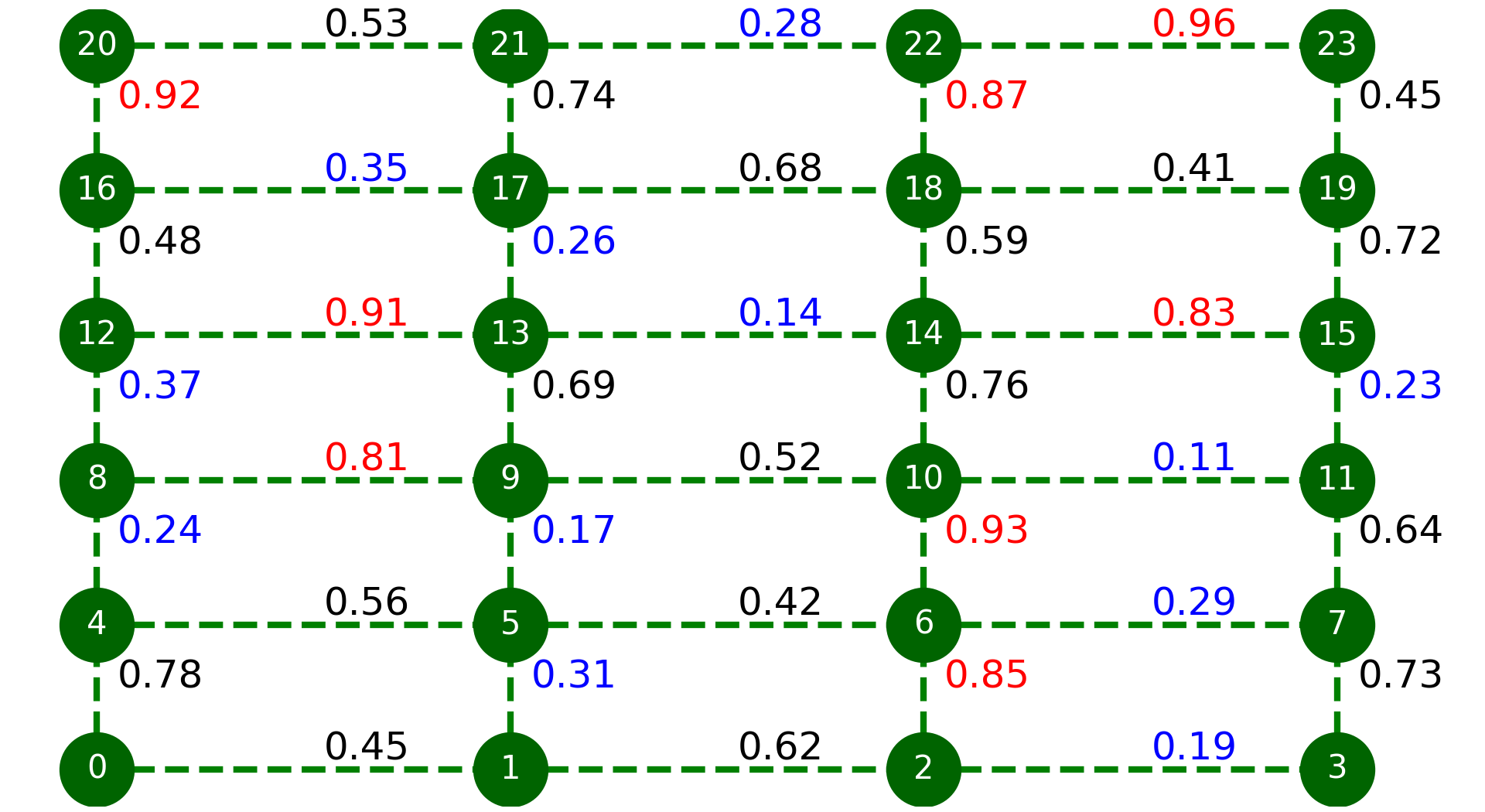}
	\caption{Abstracted sub-network representation: satellites (green) with ISLs (green-dotted), and link load (color according to load level).}
	\label{fig:network}
\end{figure}

\subsection{Reward shaping}
Defining suitable rewards is imperative to enforce the intended optimization behavior.
For CDRL, it is possible to formulate composite rewards which split feedback into global and more immediate local rewards \cite{ali_hierarchical_2020}.
At time step $t$, the global reward of an agent  can capture relevant end-to-end characteristics.
Specifically, the flow-specific latency $L(t)$, the maximum or average link utilization $U(t)$, and the resulting packet dropping rate $D(t)$ can be considered.
Each metric is weighted by a factor, denoted as $w_1$, $w_2$, and $w_3$.
However, this reward is only received after a significant time lag, denoted by $T$, due to the multi-hop propagation delays.
The global reward can thus be represented as:
\begin{equation}
    r_{t+T}^G = w_1 U(t) - w_2 L(t) - w_3 D(t)
\end{equation}
The local reward represents immediate feedback consisting of directly observable metrics in the scope of a node and its neighborhood.
Thus, it is assumed to be received in the subsequent time step $t+1$.
In this context, local link thresholds $\delta_{th}^{L}(t)$, local delays $l^L(t)$, and packet losses $d^L(t)$ due to saturated links can be considered.
As before, these metrics are weighted by factors, denoted as $w_4$, $w_5$, and $w_6$.
The local reward can thus be described by:
\begin{equation}
r_{t+1}^L = w_4 \delta_{th}^{L}(t) - w_5 l^L(t) - w_4 d^L(t)
\end{equation}
In our investigation of FD-MADRL, we specifically focus on the effectiveness of such local rewards.
For fully decentralized designs, composite rewards are typically not considered \cite{you_toward_2022, soret_q-learning_2023}.
The limited scope makes it difficult to evaluate the impact of individual actions of independent agents.
Instead of using local delays, we utilize the estimated resulting decrease in distance towards the destination, as in \cite{soret_q-learning_2023}.
We denote the value of this path reduction $\psi$.
Most importantly, to form coherent paths, loops have to be punished, but only at the nodes which are actually causing them.
In this case a punishment $-\Psi$ is received, which should be at least a magnitude higher than $\psi$.
Including a similarly large reward, so $+\Psi$, for a successful path improved training performance - even though it was not propagated to all nodes along the path.

To incentivise decentralized load balancing, a threshold-based approach proved effective.
The utilized values of $\xi$ are in the domain of $\psi$ for balanced paths.
Depending on which policy is more desired, the values for $\psi$ and $\xi$ can be varied.
$-\Xi$ punishes the usage of saturated links, which results in packet drops. As we intend to strongly discourage this behaviour, $-\Xi$ is in the domain of $-\Psi$.
The resulting local reward considered for FD-MADRL is thus the sum of:
\begin{equation}
    r'_{t+1} = \begin{cases}
    +\Psi, & \text{if goal reached}\\
    +\psi, & \text{if overall path reduction} \\
    -\psi, & \text{if overall path extension} \\
    -\Psi, & \text{if path not reduced \& loop} \\
    +\xi_1, & \text{if link load} < 0.4 \\
    -\xi_2, & \text{if } 0.4 < \text{link load} \leq 0.8 \\
    -\xi_3, & \text{if link load} > 0.8\\
    -\Xi, & \text{if link saturated}\\
    \end{cases}
\end{equation}

\section{Results}
\subsection{Simulation Environment}
To analyze the considered approaches, a specifically designed environment was created using the gym library \cite{1606.01540}.
The DQN functionality were implemented using Keras for Tensorflow \cite{chollet2015keras, tensorflow2015-whitepaper}.
The resulting grid network represents a sub-network of an SCN, e.g. a cluster in a distributed architecture \cite{roth_distributed_2022}.
For the comparisons we use a cluster of 12 satellites as well as a cluster of 24 satellites with varied link loads. The larger cluster is shown in Fig. \ref{fig:network}.
While the loads are set randomly, they comply with observed link characteristics of plausible non-uniformly distributed traffic scenarios \cite{roth_distributed_2022}.
For this performance comparison, we focus on the latency on a per-hop basis, i.e. the hop count, as well as the link loads.
Due to the limited scope of this manuscript, only the described reward design is investigated.

\subsection{Performance Evaluation}
\subsubsection{Static Path Finding}
\begin{figure}[t]
    \centering
    \includegraphics[width=.95\columnwidth]{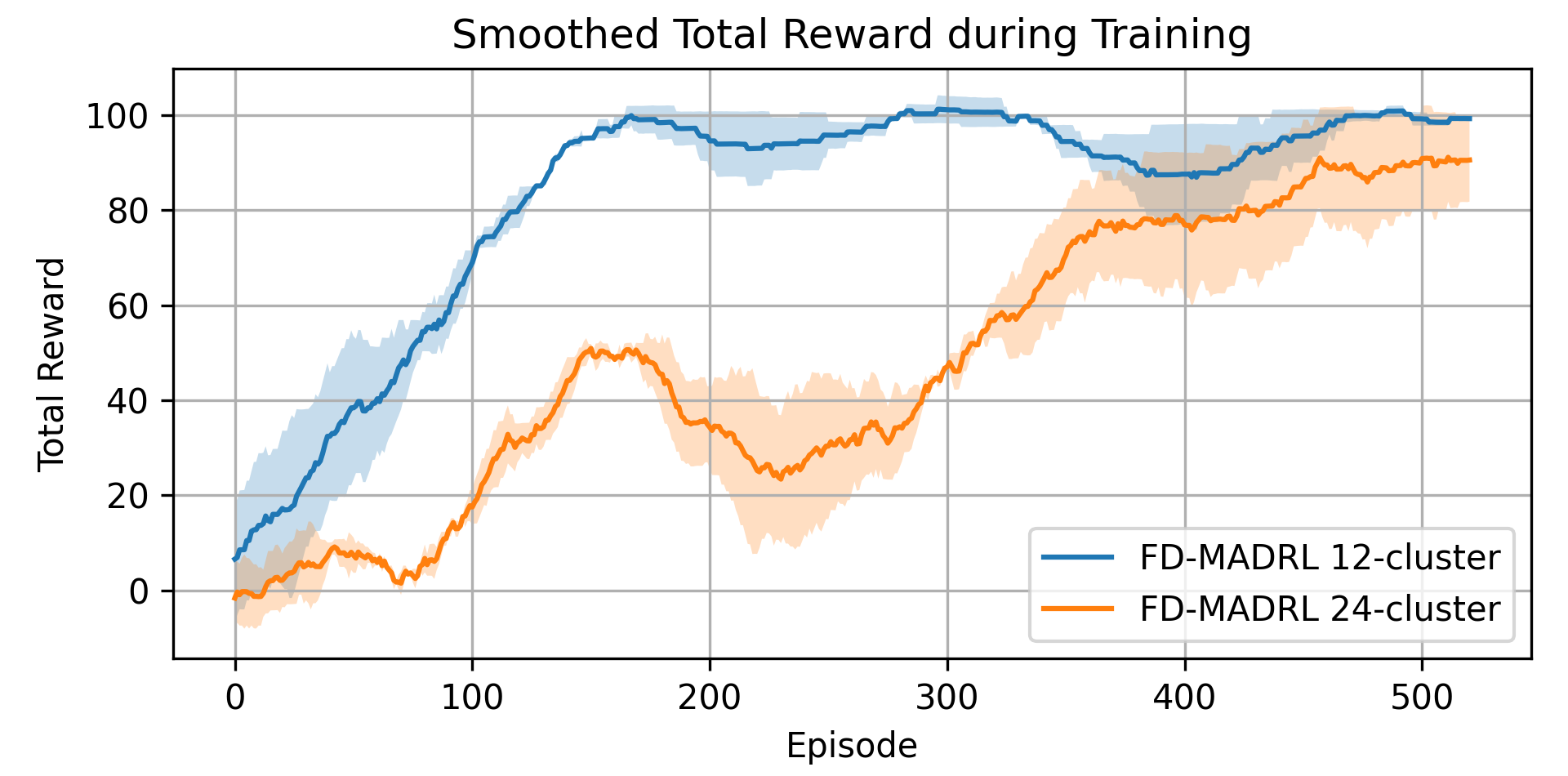}
	\caption{Smoothed rewards over training for different DQN-based approaches in static scenarios, highlighting differences in convergence.}
	\label{fig:rewards}
\end{figure}

Firstly, we investigate a static case.
The goal is to find paths with suitable latency and link load characteristics.
To this end, we compare the total rewards received during training in each episode for the 12-node and 24-node clusters to gain insights into scalability.
As shown in Fig. \ref{fig:rewards}, more than twice as many episodes are required to achieve maximum rewards consistently.
When analyzing the results, we observed that oftentimes sub-optimal decisions were made in the last hop.
For instance, in the 24-node cluster shown in Fig. \ref{fig:network}, to reach node 23 from node 4, going up to node 8 was learned, as the adjacent link has a lower load.
This however can result in a route going over node 22 which then takes the almost saturated link to 23.
As assumed, the limited scope of individual nodes can result in costly end-to-end decisions.

\begin{figure}[t]
    \centering
    \includegraphics[width=.95\columnwidth]{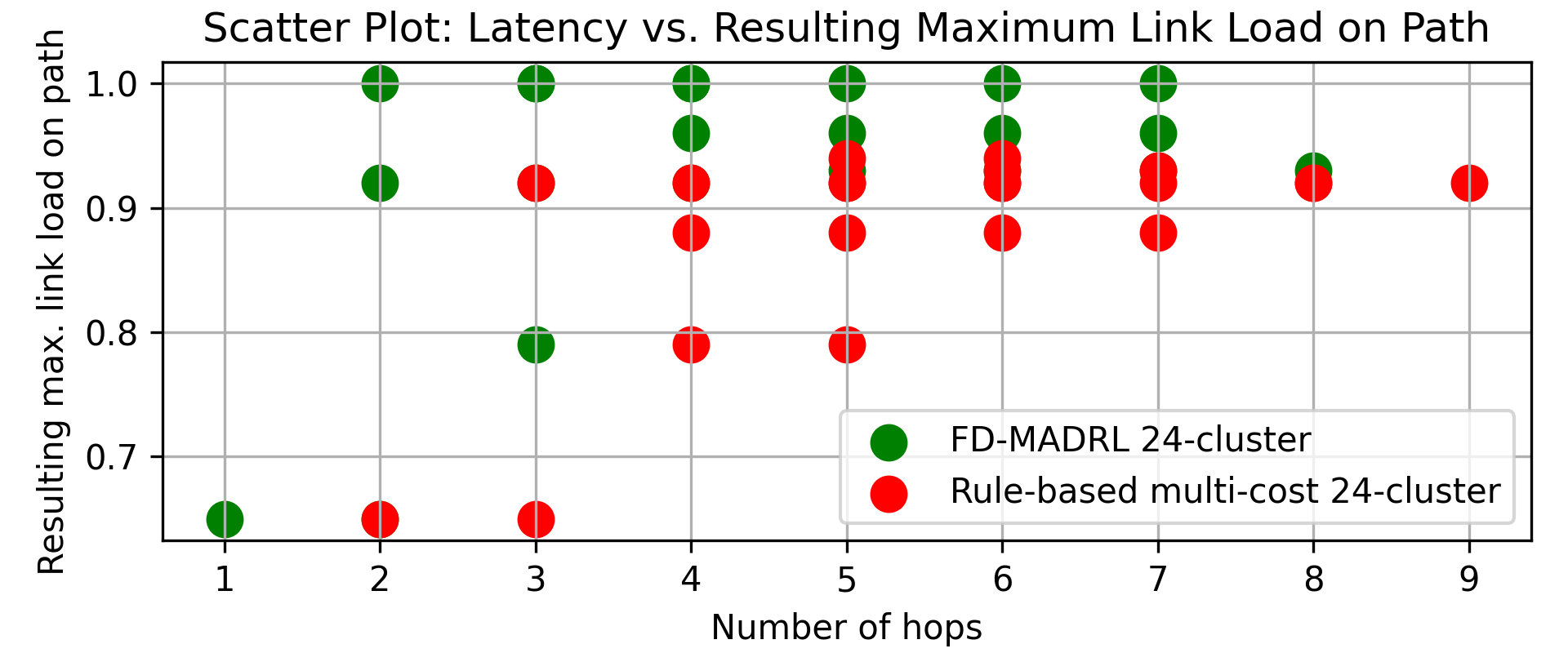}
	\caption{Comparison of FD-MADRL (in green) and a rule-based multi-cost approach (in red) in terms of latency (number of hops, on the x-axis) and the resulting maximum link load on the chosen path (y-axis). }
	\label{fig:performance}
\end{figure}

The performance of the approach in comparison to a dynamic \ac{SPF} scheme using a dynamic multi-cost approach based on Dijkstra's algorithm \cite{dijkstra_note_1959} is shown in \ref{fig:performance}. The rule-based algorithm is designed to favor short paths with minimum load, and represents a typical state-of-the-art approach \cite{roth_distributed_2022}.
Links with high link loads ($>80\%$ of capacity used) are actively avoided.
The results show that the current reward design achieves short paths: for most routes FD-MADRL requires fewer hops.
However, the approach tends to saturate links.
While this static example may not reveal significant issues, it's crucial to consider potential consequences in dynamic scenarios where a lack of foresight can result in decreased overall system performance.

\subsubsection{State Evolution: Dynamic Link Loads}
\begin{figure}[t]
    \centering
    \includegraphics[width=.95\columnwidth]{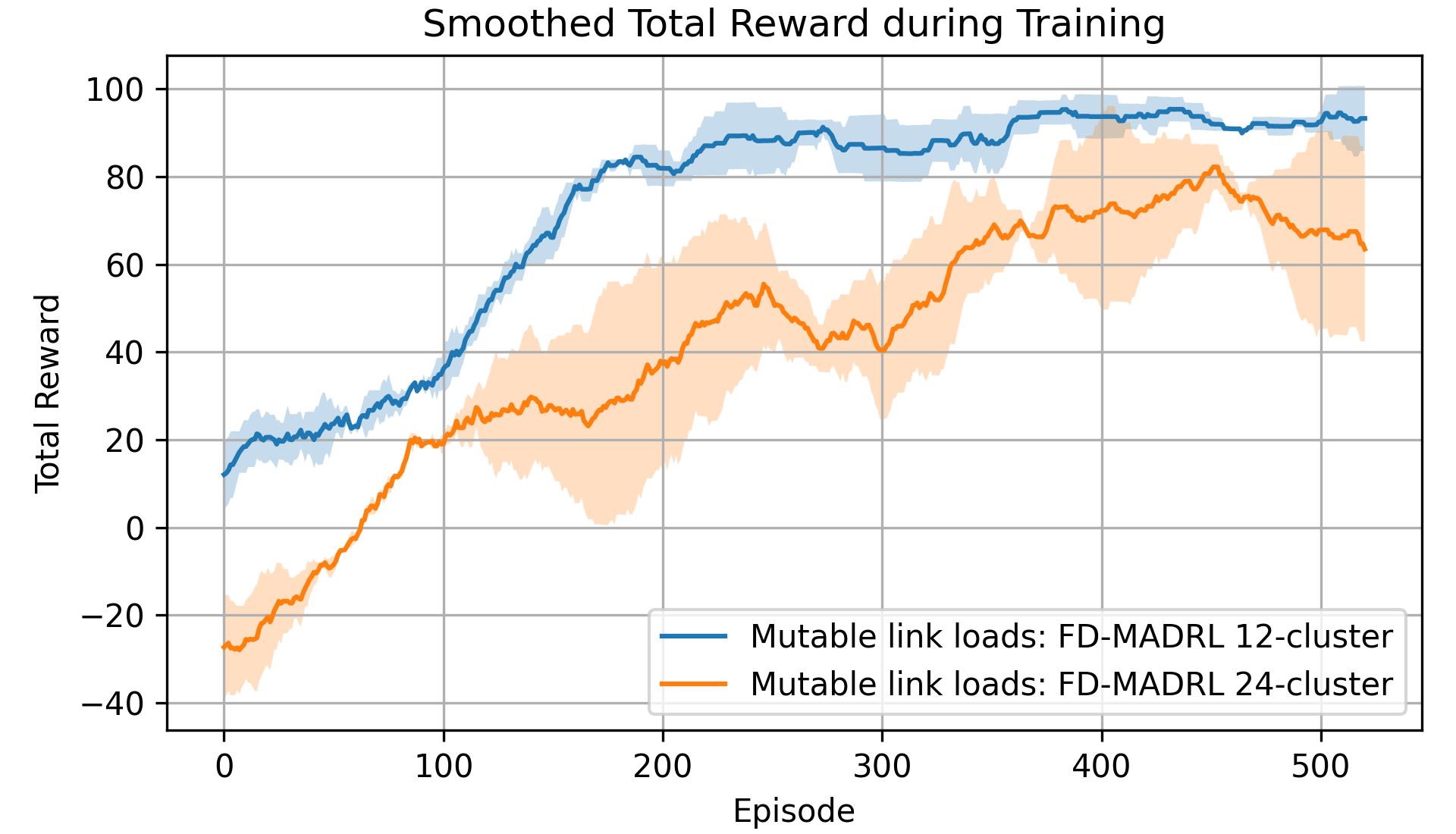}
	\caption{Smoothed rewards over training for FD-MADRL with dynamic, mutable link loads.}
	\label{fig:rewards-evo}
\end{figure}

If we reflect the impact of previous decisions in the link loads, differences in training can be observed, as shown in Fig. \ref{fig:rewards-evo}.
The idea is to test the robustness of the approach by evolving the state space according to previously made decisions.
This corresponds more to an actual SCN scenario, where routing decisions impact future states and link loads change dynamically.
In a sense, we investigate subsequent snapshots of the constellation.
To this end, we change the network to be affected by the previous action trajectory: an additional load of $20\%$ is added to each utilized link of the previous episode.
These additional loads are only episode-specific, to avoid saturation on all links.

In comparison to Fig. \ref{fig:rewards}, the achieved rewards are lower and less stable in the given training setup.
While this can be explained by higher link loads resulting in more negative rewards, the current setup is also not able to robustly respond to the more varied state space.
While a high performance can be achieved for the 12-node cluster nonetheless, the results are unstable for the 24-node cluster.
In this larger network, more scenarios can arise which lead to significant negative rewards: either by selecting saturated links or by inadvertently introducing loops.
Moreover, learning to proactively avoid potential bottlenecks is difficult with fully decentralized agents.
These results highlight the necessity for adjustments such as information exchange between nodes, or schemes specifically designed to enable cooperation between agents.

\section{Discussion}
The results show that FD-MADRL is capable of converging to correct routing policies for the considered scenario, in terms of hop count and load balancing.
However, with increasing cluster size, and state complexity, the training is significantly more difficult.
More varied \ac{QoS} requirements and network dynamics will require even more complex state representations and reward structures.
Since DRL-based approaches are expected to outperform conventional, rule-based algorithms in such complex scenarios, they are essential for the evaluation of the proposed schemes.
For FD-MADRL, we have shown that the limited scope of each agent negatively impacts the end-to-end routes, in both static and dynamic scenarios.
Moreover, with increasingly complex policies, the agents have to learn the behavior of other agents, which is counterproductive for scalability.
A CDRL-based scheme on the other hand, loses the flexibility of decentralization, introduces a single point of failure, and needs additional signalling.

\begin{figure}[t]
    \centering
    \includegraphics[width=\columnwidth]{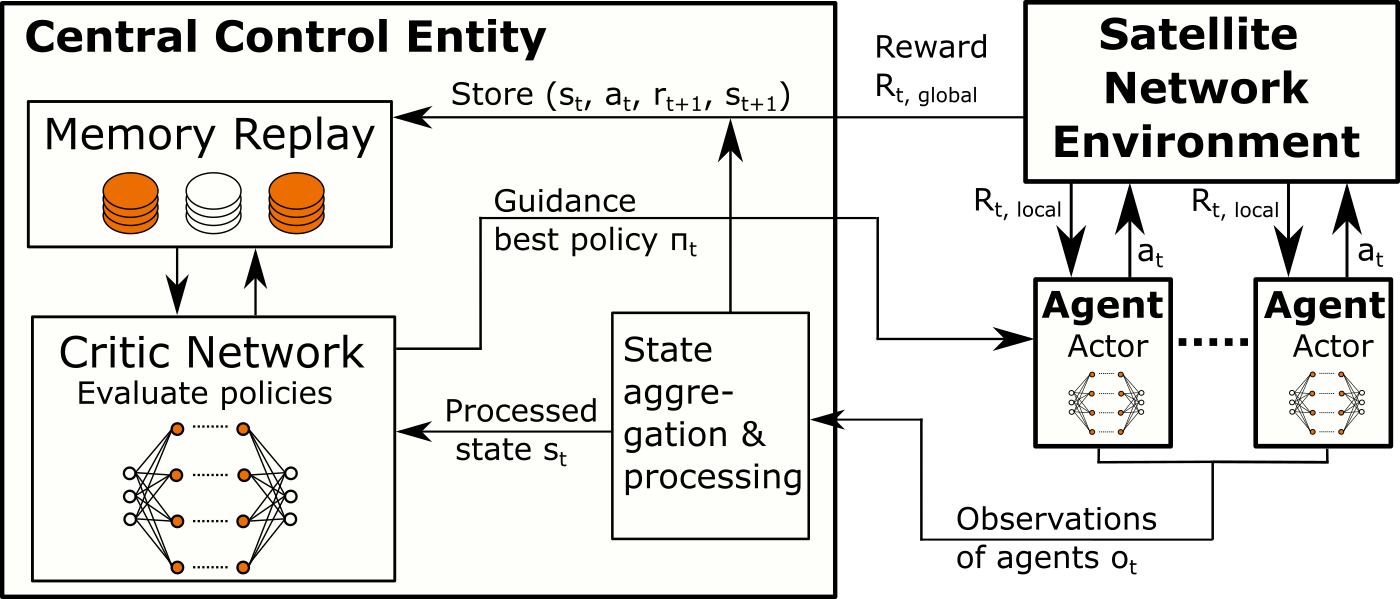}
	\caption{Proposed extended architecture: actor-critic approach using Centralized Learning and Decentralized Control (CL-DC). As in FD-MADRL actors make decisions locally, but receive central guidance from critic network.}
	\label{fig:critic}
\end{figure}

To balance these concepts, future research may consider a hybrid approach based on \ac{CL-DC}.
Such a scheme can improve coherence and end-to-end performance while maintaining the flexibility, robustness, and scalability of FD-MADRL.
In this setup, a central controller provides guidance, while decisions are made locally by the agents.
Local actions and rewards are forwarded to the central entity for learning.
The centralized guidance improves stability by facilitating coordination between agents.
Based on local conditions and constraints, the decentralized agents can make informed local decisions.
As the Q-function generally requires the same information at training and execution time, future research may also focus on policy gradient techniques for multi-agent actor-critic architectures \cite{lowe_multi-agent_2017}.
State aggregation and pre-processing may be considered to facilitate training.
This hybrid approach combining aspects from FD-MADRL and actor-critic methods is depicted in Fig. \ref{fig:critic}.

In conclusion, while DQN-based architectures represent promising solutions for routing and network control in complex SCNs, they still have practical limitations.
The investigated FD-MADRL approaches face scaling challenges, particularly in mastering complex scenarios where they are expected to surpass state-of-the-art methods.
In-depth investigations are required to fully evaluate their actual viability.

\section*{Acknowledgment}
This work has been supported by the Space for 5G \& 6G programme of the European Space Agency (ESA), activity code $5A.079$, \url{https://connectivity.esa.int/projects/aicoms}.
Responsibility for the contents of this publication rests with the authors.

\printbibliography
\addcontentsline{toc}{section}{References}

\end{document}